\documentclass[runningheads]{llncs}
\usepackage[T1]{fontenc}
\usepackage{graphicx}
\usepackage{amsmath}
\usepackage{dsfont}
\usepackage{xcolor}
\usepackage{subcaption}
\usepackage{booktabs}  % tables
\usepackage{multirow}  % tables
\usepackage{multicol}  % tables
\usepackage{siunitx} 
\usepackage{hyperref}
\usepackage{color}

\begin{document}
\title{Self-supervised Vision Transformer are Scalable Generative Models for Domain Generalization}
\titlerunning{Self-supervised Vision Transformer are Scalable Generative Models}
\author{Sebastian Doerrich \and 
Francesco Di Salvo \and
Christian Ledig}
\authorrunning{S. Doerrich et al.}
\institute{xAILab Bamberg, University of Bamberg, Germany
\email{sebastian.doerrich@uni-bamberg.de}}
\maketitle              % typeset the header of the contribution
\begin{abstract}
Despite notable advancements, the integration of deep learning (DL) techniques into impactful clinical applications, particularly in the realm of digital histopathology, has been hindered by challenges associated with achieving robust generalization across diverse imaging domains and characteristics. Traditional mitigation strategies in this field such as data augmentation and stain color normalization have proven insufficient in addressing this limitation, necessitating the exploration of alternative methodologies. To this end, we propose a novel generative method for domain generalization in histopathology images. Our method employs a generative, self-supervised Vision Transformer to dynamically extract characteristics of image patches and seamlessly infuse them into the original images, thereby creating novel, synthetic images with diverse attributes. By enriching the dataset with such synthesized images, we aim to enhance its holistic nature, facilitating improved generalization of DL models to unseen domains. Extensive experiments conducted on two distinct histopathology datasets demonstrate the effectiveness of our proposed approach, outperforming the state of the art substantially, on the \textsc{Camelyon17-wilds} challenge dataset (+2\%) and on a second epithelium-stroma dataset (+26\%). Furthermore, we emphasize our method's ability to readily scale with increasingly available unlabeled data samples and more complex, higher parametric architectures. Source code is available at \href{https://github.com/sdoerrich97/vits-are-generative-models}{github.com/sdoerrich97/vits-are-generative-models} .

\keywords{domain generalization \and self-supervised learning \and feature orthogonalization \and generative image synthesis.}
\end{abstract}
\section{Introduction}
\label{sec:introduction}
Deep learning (DL) has had a significant impact on a broad range of domains ranging from image classification to natural language processing \cite{Wang2023}. Nevertheless, its incorporation into routinely used medical image analysis has progressed comparatively slow \cite{Stacke2021}, mainly due to difficulties in achieving robust generalization across diverse imaging domains. This challenge is particularly pronounced in digital histopathology, where variations in coloring agents and staining protocols for histological specimens exacerbate domain disparity \cite{Moscalu2023}.
Traditional approaches to address these generalizability challenges in digital histopathology typically involve data augmentation or stain color normalization \cite{Chang2021}. Data augmentation techniques manipulate aspects of color \cite{Liu2017DetectingCM}, apply stain-specific channel-wise augmentation \cite{Tellez2018}, or incorporate stain colors of unseen domains into the training data \cite{Chang2021}. Alternatively, stain color normalization aligns images' color patterns using target domain information \cite{Reinhard2001,Macenko2009,Vahadane2016StructurePreservingCN}. However, these methods often require access to target samples during training or struggle with adapting to new domains and unseen stain colors.
To overcome these limitations, Lafarge et al. \cite{Lafarge2017} investigate the use of Domain Adversarial Neural Networks (DANNs) to enhance cross-domain performance. Conversely, Nguyen et al. \cite{nguyen2023contrimix} propose ContriMix, which aims to improve domain generalization by augmenting the diversity of the source domain with synthetic images. This is achieved by initially separating biological content from technical variations and subsequently combining them to form new anatomy-characteristic combinations. However, ContriMix's dependence on convolutional encoders restricts the diversity of its synthetic images, as it allows for the extraction of only a single characteristic tensor per image.
\begin{figure}[t]
    \centering
    \includegraphics[width=0.83\linewidth]{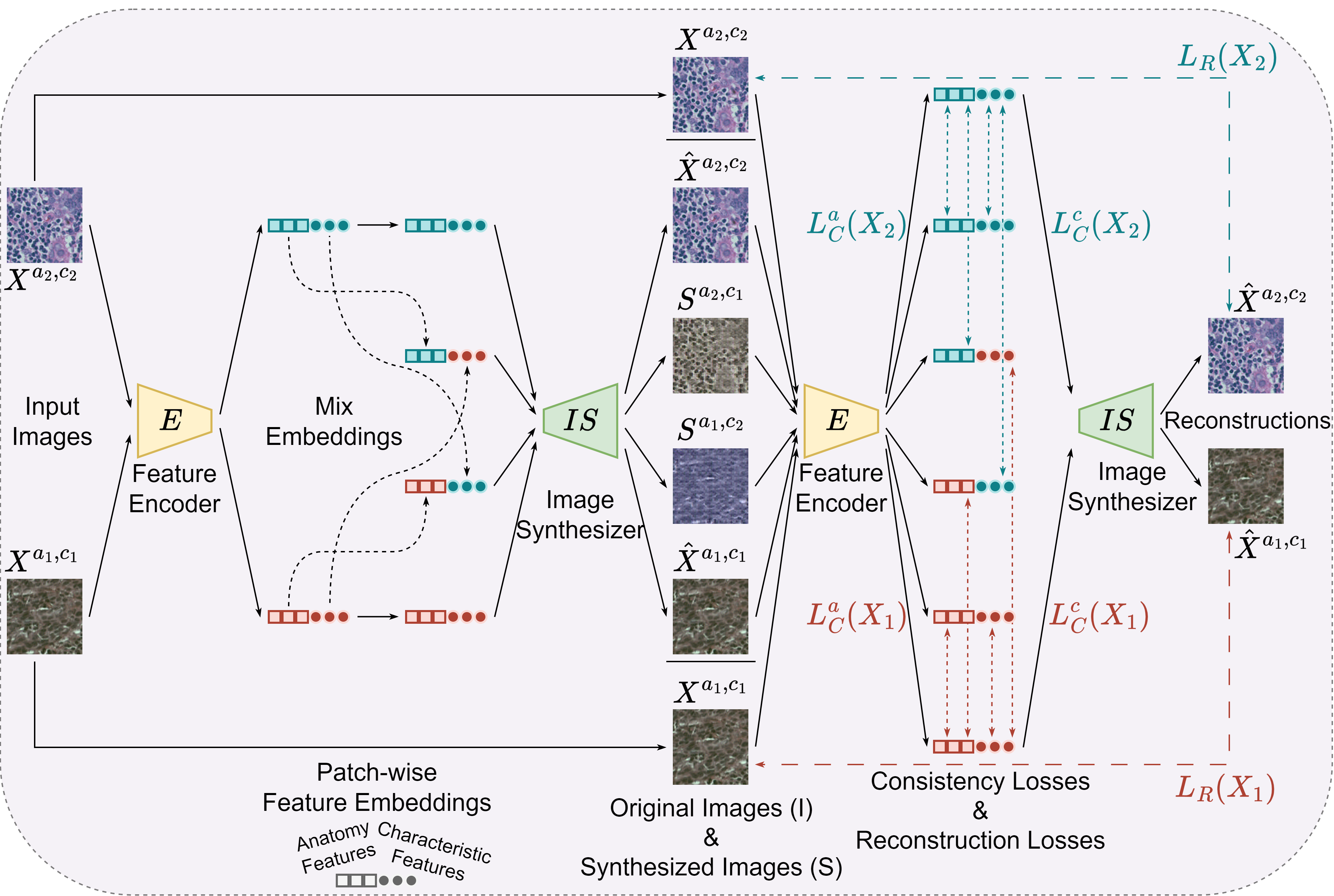}
    \caption{Schematic Visualization of our self-supervised generative approach. A single ViT encoder ($E$) is used to separate anatomy from image-characteristic features of distinct images which are subsequently intermixed among each other and processed by an image synthesizer ($IS$) to generate synthetic images.}
    \label{fig:method}
\end{figure}
In this work, we focus on those limitations and present a novel generative domain generalization (DG) method for histopathology images. Employing a self-supervised Vision Transformer (ViT), we generate synthetic images with diverse combinations of anatomy and image characteristics, enriching the holistic nature of the dataset without requiring any domain information. This allows DL models trained on the extended dataset to adapt to unseen domains more effectively. To prove this, we evaluate our method in extensive experiments against the current state of the art on two distinct benchmark datasets for domain generalization in histopathology.

Our main contributions are:
\begin{itemize}
    \item We present a novel self-supervised generative domain generalization method for histopathology.
    \item We generate synthetic images with unseen combinations of anatomy and image characteristics.
    \item We extensively evaluate our method on two histopathology benchmark datasets and outperform the state of the art by a large margin.
    \item We assess our method's ability to scale effectively with growing availability of unlabeled data samples and the adoption of deeper architectures.
\end{itemize}
\section{Method}
\label{sec:method}
Our method is a self-supervised generative approach that employs feature orthogonalization to generate synthetic images. Using a single ViT encoder ($E$), we encode an image patch-wise and split the resulting embeddings, with one half preserving anatomy and the other half storing characteristic features for each patch. These feature vectors are then mixed across different input images and fed into an image synthesizer ($IS$) to create synthetic images representing new anatomy-characteristic pairs. See \figurename~\ref{fig:method} for an illustration of this process.
\subsection{Feature Orthogonalization and Image Synthesis}
\label{subsec:orthogonalization_and_synthesis}
Taking inspiration from ViT principles \cite{dosovitskiy2021}, we first partition images $x_i$ with $x_i \in \mathds{R}^{C \times H \times W}$, where $C$, $H$, and $W$ are the number of channels, height, and width of the image, respectively, into non-overlapping patches. This results in $\tilde{x}_i \in \mathds{R}^{P \times C \times PS \times PS}$, where $P$ denotes the number of patches and $PS$ the patch size. These patches are processed by the encoder $E$ to extract feature embeddings $z_i$ for each image. Let $z_i = E\left(\tilde{x}_i\right) \in \mathds{R}^{P \times L}$, where $L$ denotes the encoder's latent dimension, we extract the anatomical ($z_{i}^{a} \in \mathds{R}^{P \times L / 2}$) and characteristic ($z_{i}^{c} \in \mathds{R}^{P \times L / 2}$) feature vectors by splitting $z_i$ along $L$. To reconstruct the original images $\hat{x}_i$, the image synthesizer $IS$ reshapes the feature vectors into matrices $Z_{i}^{a} \in \mathds{R}^{P \times C \times PS \times V}$ and $Z_{i}^{c} \in \mathds{R}^{P \times C \times V \times PS}$, where $V$ is the hidden dimension, before applying the dot-product of both feature matrices along $V$ to restore $\hat{x}_i$.
\begin{equation}
    \hat{x}_i = IS\left(z_{i}^{a}, z_{i}^{c}\right) = Z_{i}^{a} \cdot Z_{i}^{c}, \qquad \text{with}\; \hat{x}_i \in \mathds{R}^{P \times C \times PS \times PS} \longleftrightarrow \mathds{R}^{C \times H \times W}
\end{equation}
Conversely, to generate synthetic images $s_i$ with diverse anatomy-characteristics combinations, we combine the anatomical feature embeddings $z_{i}^{a}$ of each sample $x_i$ in batch $b$ with $M$ characteristic feature embeddings. These are each extracted from a single patch of another sample $x_m$ within the same batch (\( m \in 1, \dots, M \)). This patch, and thereby its corresponding characteristic embedding $z_{m,p}^{c}$ are chosen uniformly at random from each sample $x_m$. Note that we do not use the entire $z_{m}^{c}$ since using the characteristics of a single patch yields substantially more diverse synthetic images. These combinations ($z_{i}^{a}$, $z_{m,p}^{c}$) are then passed through $IS$ to create the synthetic images $s_i$, preserving the original anatomy but with severely altered characteristics. This process enables the extraction of fine-grained characteristics, resulting in a diverse range of synthetic images $s_i$.
\subsection{Feature Consistency and Self-Reconstruction}
\label{subsec:losses}
To guide the feature orthogonalization and synthetic image generation, we employ three distinct mean squared error (MSE) loss terms, namely anatomical consistency $L_{C}^{a}$, characteristic consistency $L_{C}^{c}$ and self-reconstruction $L_{R}$. The anatomical consistency $L_{C}^{a}$ for batch $b$ with $N$ training samples and $M$ number of anatomy-characteristic mixes: 
\begin{equation}
    \begin{gathered}
    L_{C}^{a} = \frac{1}{NM} \sum_{i = 1}^{N} \sum_{m = 1}^{M} \left|\left| z_{i}^{a} - z_{s}^{a} \right|\right|_{2}^{2}
    \\
    \text{with} \quad z_{i}^{a} = E\left(x_i\right)^{P \times [1\,:\,L/2]} \quad \text{and} \quad z_{s}^{a} = E\left(IS\left(z_{i}^{a}, z_{m, p}^{c}\right)\right)^{P \times [1\,:\,L/2]}
    \end{gathered}
\end{equation}
where $z_{m, p}^{c}$ being the characteristic embedding of a randomly chosen patch $p$ of sample $x_m$, promotes consistency between the anatomy extracted from the original images $x_{i}$ and the corresponding synthetic images $s_{i}$. In addition, the characteristic consistency $L_{C}^{c}$ for batch $b$ with $N$ training samples and $M$ number of anatomy-characteristic mixes:
\begin{equation}
    \begin{gathered}
    \quad L_{C}^{c} = \frac{1}{NMP} \sum_{i = 1}^{N} \sum_{m = 1}^{M} \sum_{q = 1}^{P} \left|\left| z_{m,p}^{c} - z_{s,q}^{c} \right|\right|_{2}^{2}
    \\
    \text{with} \quad z_{s,q}^{c} = E\left(IS\left(z_{i}^{a}, z_{m, p}^{c}\right)\right) \text{ at patch } q \in P \quad \text{and} \quad z_{s,q}^{c} \in \mathds{R}^{1 \times L / 2}
    \end{gathered}
\end{equation}
aligns the characteristics of the synthetic images $s_{i}$ with the characteristic $z_{m, p}^{c}$ used to create these synthetic images. Lastly, the self-reconstruction loss $L_{R}$:
\begin{equation}
    L_{R} = \frac{1}{N} \left|\left| x_{i} - IS\left(z_{i}^{a}, z_{i}^{c}\right) \right|\right|_{2}^{2}
\end{equation}
aims to ensure that the self-reconstructed images closely resemble the original ones. Thereby, the combined loss across a set of mini-batches with $b \in 1, \dots, B$ can be written as:
\begin{equation}
    L = \frac{1}{B} \sum_{b = 1}^{B} \lambda_a L_{C}^{a} + \lambda_c L_{C}^{c} + \lambda_r L_{R}
\end{equation}
with $\lambda_a$, $\lambda_c$, $\lambda_r$ being weights to adjust the influence of each loss during training.
\subsection{Training}
\label{subsec:training}
The encoder is trained independently for each dataset adhering to the objective described above. This fully self-supervised approach allows us to incorporate labeled or unlabeled samples for the anatomical area of interest and facilitates dynamic transfer to additional tasks without retraining. For the ViT encoder $E$, we opt for the ViT-B/16 backbone, which operates on $224 \times 224$ pixel images, splitting them into $16 \times 16$ pixel patches and encoding each patch into a $768$-dimensional vector. Following \cite{nguyen2023contrimix}, we use $4$ mixes (number of combinations $M$ of anatomy and characteristics to get synthetic images) per batch. We set $\lambda_a = \lambda_c = \lambda_r = 1$ and train the encoder for $50$ epochs with a batch size of $64$, utilizing the AdamW optimizer \cite{Loshchilov2017DecoupledWD} with a learning rate of $0.001$, and a cosine annealing learning rate scheduler \cite{Loshchilov2016SGDRSG} with a single cycle.
\section{Experiments and Results}
\label{sec:experimentsAndResults}
\begin{figure}
    \hfil
    \begin{subfigure}[b]{0.41\textwidth}
        \includegraphics[width=\textwidth]{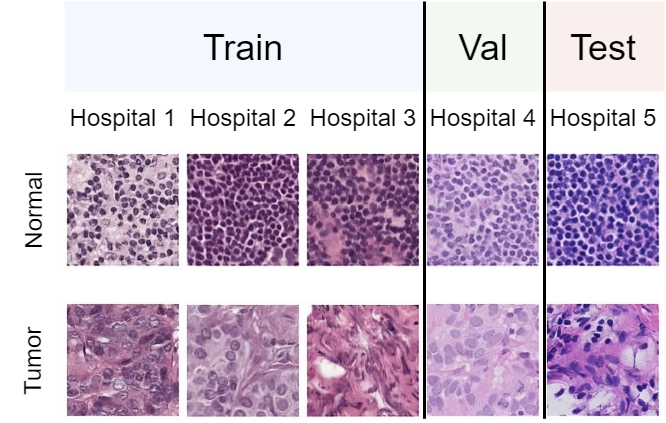}
        \caption{\textsc{Camelyon17-wilds}}
        \label{fig:camelyon17}
    \end{subfigure}
    \hfil
    \begin{subfigure}[b]{0.41\textwidth}
        \includegraphics[width=\textwidth]{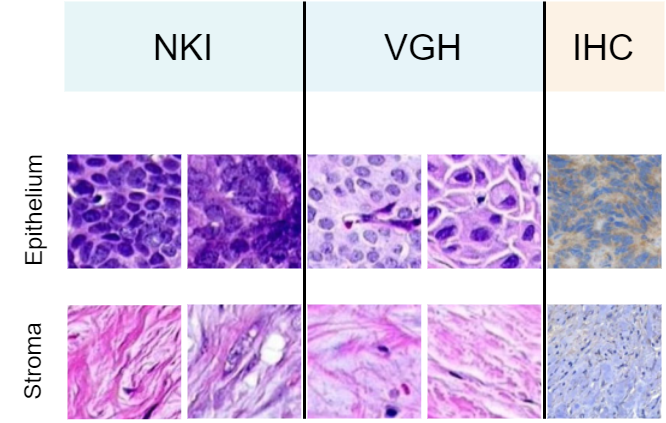}
        \caption{Epithelium-Stroma}
        \label{fig:epistr}
    \end{subfigure}
    \hfil
    \caption{Examples from the histopathology datasets used for evaluating domain generalization. Left: \textsc{Camelyon17-wilds} for which the domains are hospitals. Right: Combined epithelium-stroma dataset for which the domains are datasets.}
    \label{fig:dataset}
\end{figure}

We assess the domain generalization ability of our method on two histopathology datasets. The first is the \textsc{Camelyon17-wilds} challenge dataset \cite{wilds2021,wilds2022}, focusing on tumor identification across various hospitals. It comprises $96 \times 96$ image patches from lymph node whole-slide images, with labels indicating tumor presence in the central $32 \times 32$ region. We use the same training (302,436 samples), validation (34,904), and test (85,054) splits as the original publication \cite{wilds2021}.
For the second dataset, we aggregate three public histopathology datasets: NKI \cite{Beck2011}, VGH \cite{Beck2011}, and IHC \cite{Linder2012}, focusing on epithelium-stroma classification. The NKI (8,337 samples) and VGH (5,920) datasets comprise H\&E stained breast cancer tissue images, while the IHC dataset (1,376) consists of IHC-stained colorectal cancer tissue images. Following \cite{Huang2017}, we alternate between NKI and VGH as the train/validation set, but maintain IHC as the fixed test set due to its distinct coloration. This allows us to mimic a similar generalization challenge as presented in \textsc{Camelyon17-wilds}, where both the validation and test set comprise out-of-distribution (OOD) samples. In order to fully utilize our ViT encoder's abilities, both benchmark datasets are standardized to $224 \times 224$ images using bicubic interpolation. Examples for each dataset are illustrated in \figurename~\ref{fig:dataset}.
\subsection{Qualitative Evaluation}
\label{subsec:qualitative}
\begin{figure}
    \centering
    \includegraphics[width=0.61\linewidth]{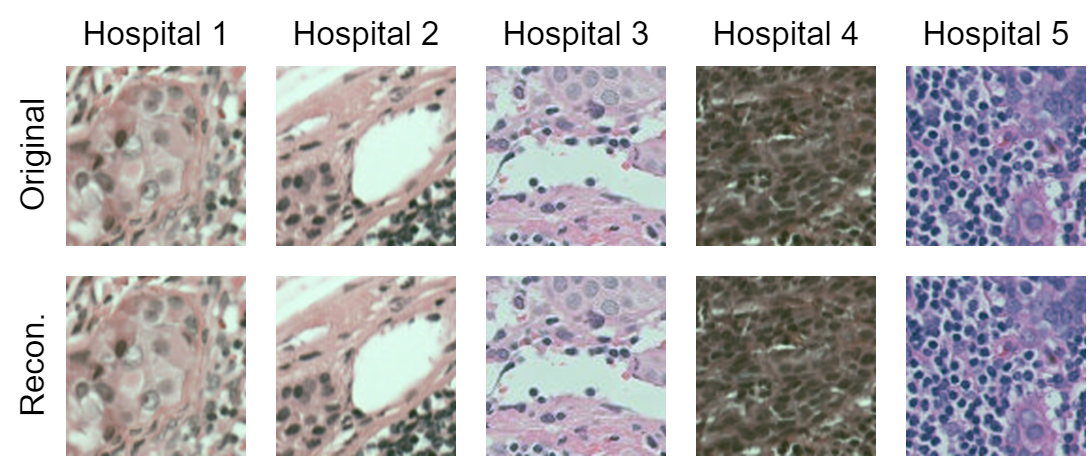}
    \caption{Qualitative evaluation of our method's reconstruction capability on the \textsc{Camelyon17-wilds} dataset.}
    \label{fig:reconstruction}
\end{figure}

We qualitatively evaluate our method by training it on the \textsc{Camelyon17-wilds} dataset and assessing the image quality of the image synthesizer's reconstructions (no mixing). For the training set, we achieve an average Peak Signal-to-Noise Ratio (PSNR) of \qty{46}{\dB}, for the OOD validation set of \qty{46}{\dB} and for the OOD test set of \qty{40}{\dB}. These results demonstrate the model's capability to successfully encode image information while retaining a holistic understanding in order to generalize to unseen domains. \figurename~\ref{fig:reconstruction} illustrates this qualitatively for 5 distinct samples from each hospital and dataset split.

\begin{figure}[!h]
    \centering
    \includegraphics[width=0.9\linewidth]{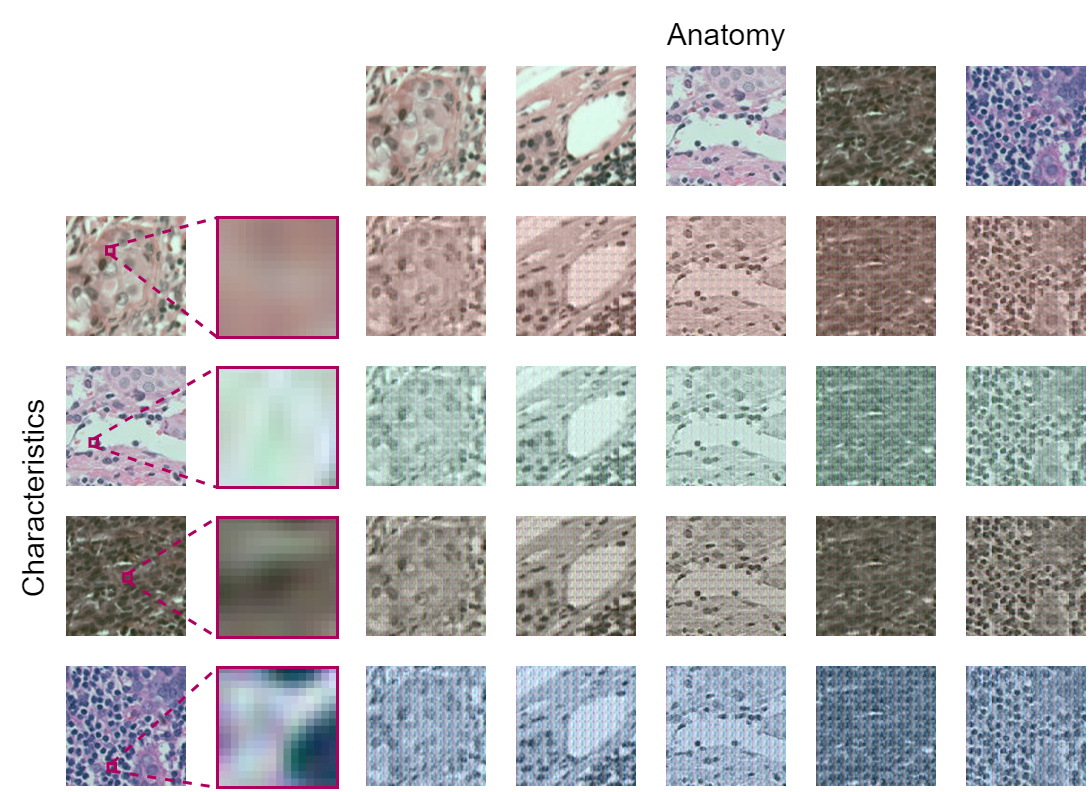}
    \caption{Qualitative evaluation of the method's generative capabilities on the \textsc{Camelyon17-wilds} dataset by means of synthetic images created through its anatomy-characteristics intermixing.}
    \label{fig:mixing}
\end{figure}

\noindent We also assess the image quality of synthetic images, which exhibit the same anatomy but varied characteristics, generated by our image synthesizer. \figurename~\ref{fig:mixing} demonstrates this process, utilizing randomly extracted patch characteristics for each row. Although our method's patch-wise image reconstruction may produce slight grid artifacts, the synthetic images accurately preserve the original anatomy while displaying uniformly the applied characteristics from the extracted patch. This approach facilitates the generation of a diverse array of samples by altering colorization while maintaining diagnostically relevant anatomy.
\subsection{Disease Classification}
\label{subsec:classification}
To evaluate our method's suitability for improving domain generalization, we employ our stand-alone encoder to generate additional synthetic images with mixed anatomy and characteristics, augmenting the training set diversity on the fly. These synthetic images, alongside the originals, are afterward fed into a subsequent classifier allowing it to learn from a more diverse set of samples, thereby generalizing better to unseen images. For the classifier, we use the same DenseNet-121 architecture \cite{Huang2016DenselyCC} used by the baseline methods in WILDS \cite{wilds2021}. We evaluate our method on the class-balanced \textsc{Camelyon17-wilds} validation and test sets against the top-performing methods from the WILDS leaderboard\footnote{\texttt{\url{https://wilds.stanford.edu/leaderboard/\#camelyon17}}}, which utilize the same classifier. The results shown in \tablename~\ref{tab:camelyon17-classification} reveal our method's superior accuracy on both sets, setting a new state-of-the-art standard.
\begin{table}
\caption{Accuracy in \% on the validation and test set of \textsc{Camelyon17-wilds}.}
\label{tab:camelyon17-classification}
    \centering
    \begin{tabular}{l c c c}
        \toprule
        \textbf{Methods} &  & \textbf{Val (OOD)} & \textbf{Test (OOD)} \\
        \midrule      
        ERM\cite{wilds2021} & & 85.80 &  70.80 \\
        LISA\cite{Yao2022} & & 81.80 & 77.10 \\
        ERM with targeted augmentation\cite{gao2022outofdistribution} & & 92.70 & 92.10 \\
        MBDG\cite{robey2021modelbased} & & 88.10 & 93.30 \\
        ContriMix\cite{nguyen2023contrimix} & & 91.90 & 94.60 \\
        \midrule    
        Ours & & \textbf{94.16} & \textbf{95.44} \\
        \bottomrule
    \end{tabular}
\end{table}

\noindent We further evaluate our method for the binary classification task of the adapted epithelium-stroma dataset. For this, we train it once on NKI and evaluate it for VGH (val) and IHC (test), as well as train it on VGH and evaluate it for NKI (val) and IHC (test), respectively. We compare the performance against the three domain adaptation methods referenced in \cite{Huang2017}. The consistent performance of our method across these evaluations, as presented in \tablename~\ref{tab:epistr-classification}, confirms its strong generalizability potential, clearly outperforming the state of the art.
\begin{table}
\caption{Accuracy in \% on the epithelium-stroma dataset.}
\label{tab:epistr-classification}
    \centering
    \begin{tabular}{l c c c c c c}
        \toprule
        \multirow{2.5}{*}{\textbf{Methods}} &  & \multicolumn{2}{c}{\textbf{Training NKI}} &  & \multicolumn{2}{c}{\textbf{Training VGH}} \\
        \cmidrule(r){3-4}\cmidrule(l){5-7}
        &  & VGH & IHC & & NKI & IHC \\
        \midrule
        DLID\cite{Chopra2013DLIDDL} & & 75.70 & 56.39 &  & 86.70 & 57.36 \\
        DDA\cite{Glorot2011DomainAF} & & 77.50 & 73.17 &  & 81.00 & 52.46 \\
        CKA\cite{Huang2017} & & 77.75 & 73.19 &  & 80.17 & 59.44 \\
        \midrule     
        Ours & & \textbf{93.72} & \textbf{85.39} &  & \textbf{88.47} & \textbf{86.12} \\
        \bottomrule
    \end{tabular}
\end{table}
\subsection{Scalability Potential}
\label{subsec:scalability}
Finally, we investigate the scalability potential of our method to enhance its reconstruction and image synthesis capabilities. First, we exploit the label-free nature of our encoder ($E$), enabling the inclusion of unlabeled samples alongside labeled ones during training. This approach allows $E$ to learn from a larger more diverse dataset. To evaluate this, we augment our training data with an additional 302,436 (same amount as labeled training samples) randomly selected samples from the 1,799,247 unlabeled samples available in the \textsc{Camelyon17-wilds} dataset \cite{wilds2022}. Through this augmentation, our encoder achieves improved reconstruction performance compared to the base model: \qty{49}{\dB} versus \qty{46}{\dB} for the training set, \qty{49}{\dB} versus \qty{46}{\dB} for the validation set, and \qty{44}{\dB} versus \qty{40}{\dB} for the test set.
Furthermore, leveraging a Vision Transformer (ViT) backbone allows us to readily increase model capacity by replacing the ViT-B/16 backbone (86M parameters) with the deeper and more sophisticated ViT-L/16 (322M parameters). Notably, we extend the embedding dimension from 768 to 1,056 to accommodate the requirements of our image synthesizer's matrix multiplication. Training the adapted ViT-L/16 backbone for 10 epochs on \textsc{Camelyon17-wilds} already yields enhanced results, with a reconstruction performance of \qty{49}{\dB} versus \qty{46}{\dB} for the training set, \qty{49}{\dB} versus \qty{46}{\dB} for the validation set, and \qty{42}{\dB} versus \qty{40}{\dB} for the test set.
These findings demonstrate that both scaling approaches result in superior performance compared to the base method, underscoring the method's scalability potential in terms of utilizing unlabeled samples and adopting more sophisticated network architectures.
\section{Discussion and Conclusion}
In this work, we introduce a novel self-supervised, generative method for domain generalization. By employing the power of a Vision Transformer encoder, we successfully generate synthetic images featuring diverse combinations of anatomy and image characteristics in a self-supervised fashion. This approach enriches the representativeness of the dataset without necessitating any domain-specific information, thereby enabling more effective adaptation to previously unseen domains. Through quantitative experimentation on two distinct histopathology datasets, we demonstrate the efficacy of our method. Our qualitative assessment emphasizes the model's proficiency in encoding image data and its capacity to generalize across domains. Moreover, the synthetic images generated by our method faithfully preserve original anatomical details while augmenting dataset diversity. Furthermore, by enabling the utilization of unlabeled samples or the adoption of more sophisticated ViT backbone architectures, our method demonstrates scalability potential, exhibiting improved reconstruction performance and adaptability. We believe that our method's flexibility should allow its application across various modalities for addressing generalization challenges not only in histopathology but also in other applications.
\begin{credits}
\subsubsection{\ackname} This study was funded through the Hightech Agenda Bayern (HTA) of the Free State of Bavaria, Germany.
\end{credits}
\bibliographystyle{splncs04}
\bibliography{Paper-0740}
\end{document}